\documentclass{Interspeech}
\usepackage{subfigure} 
\usepackage{subcaption}
\usepackage{amsmath}

\interspeechcameraready

\title{Speech Enhancement with Dual-path Multi-Channel Linear Prediction Filter and Multi-norm Beamforming\thanks{$^{*}$ Corresponding author}}

\author[affiliation={1}]{Chengyuan}{Qin}
\author[affiliation={1,*}]{Wenmeng}{Xiong}
\author[affiliation={2}]{Jing}{Zhou}
\author[affiliation={1}]{Maoshen}{Jia}
\author[affiliation={1}]{Changchun}{Bao}

\affiliation{Speech and Audio Signal Processing Laboratory}{Beijing University of Technology}{China}
\affiliation{Institute of AI (TeleAI)}{China telecom}{China}

\email{qinchengyuan@emails.bjut.edu.cn, wenmeng.xiong@bjut.edu.cn, zhouj100@chinatelecom.cn, jiamaoshen@bjut.edu.cn, baochch@bjut.edu.cn}

\usepackage{comment}

\begin{document}

\maketitle

\begin{abstract}

In this paper, we propose a speech enhancement method using dual-path Multi-Channel Linear Prediction (MCLP) filters and multi-norm beamforming. Specifically, the MCLP part in the proposed method is designed with dual-path filters in both time and frequency dimensions. For the beamforming part, we minimize the power of the microphone array output as well as the $l_1$ norm of the denoised signals while preserving source signals from the target directions. An efficient method to select the prediction orders in the dual-path filters is also proposed, which is robust for signals with different reverberation time ($T_{60}$) values and can be applied to other MCLP-based methods. Evaluations  demonstrate that our proposed method outperforms the baseline methods for speech enhancement, particularly in high reverberation scenarios.

\end{abstract}

\section{Introduction}

Speech enhancement technique is of great importance for numerous applications such as automatic speech recognition, human-machine interaction, and smart home devices\cite{1,2,3}. Conventional speech enhancement methods, including spectral subtraction \cite{4}, wiener filtering \cite{5}, and subspace-based methods \cite{6}, have been widely investigated and proven to be effective. However, their performance degrades severely  in real-world applications where noise and reverberation are present. 

To address the challenge of reverberation, numerous methods have been proposed based on the MCLP filter \cite{7}\cite{8}. In \cite{9}, the Generalized Weight Prediction Error (GWPE) method reduces the late reverberation  by minimizing the temporal Hadamard-Fischer (HF) mutual correlation of speech signals. In addition to dereverberation, methods based on beamforming \cite{10} have been proven effective for denoising. Plenty of work \cite{11,12,13} has been proposed for the integration of the MCLP-based methods and the beamforming-based methods for simultaneous dereverberation and denoising. For example, in \cite{11,12}, Minimum Variance Distortionless Response (MVDR) beamforming and WPE are employed in a cascade framework. 
In \cite{13} and \cite{14}, the MVDR beamforming and Minimum-Power Distortionless Response (MPDR) beamforming are combined with WPE in a unified joint optimization problem, respectively.

Traditional MCLP-based speech enhancement methods remove late reverberations by estimating temporal filters at each frequency bin of the time-frequency (TF) domain microphone signals. However, numerous deep-learning based speech enhancement techniques, including DPRNN \cite{15}, FullSubNet \cite{16}, TF-GridNet \cite{17}, SpatialNet \cite{18}, and DasFormer \cite{19}, have demonstrated significantly superior performance using dual-path cross-narrow band architectures. In these work, frequential dependencies as well as temporal correlations have been leveraged for a more comprehensive modeling by neural networks to learn the mappings between the inputs and the learning targets.

In this paper, we propose a speech enhancement method using dual-path MCLP filters and multi-norm beamforming. Specifically, for the MCLP part, exploiting both temporal correlations and frequential dependencies of signals in TF domain, dual-path filters in both temporal and frequency dimensions are designed to remove the late reverberations comprehensively. The $l_1$ norm constraint is incorporated into the cost functions of both dual-path MCLP and multi-norm beamforming because information-bearing signals, such as speech signals, exhibit sparsity in their short-time Fourier Transform (STFT) coefficients.  
Additionally, we propose an efficient method to determine the prediction orders in the temporal and frequential filters of the MCLP part. By calculating the Pearson correlation coefficients between signals on a single microphone as a function of different time or frequency lags and selecting an appropriate threshold, the corresponding time or frequency lag can be considered as the approximate optimal prediction orders. This method
 is robust for signals with varying reverberation time ($T_{60}$) values and can be applied to other MCLP-based methods. Experiments demonstrate the advantages of our proposed method over the baseline methods. 

\section{Preliminaries}
\subsection{Microphone array signal model}
Considering that $Q$ far-field wideband acoustic sources impinge on $M$ microphones in a noisy and reverberant room. The signals received at the microphone array in the TF domain are approximately formulated as a $K$ order convolution between the STFT of the room impulse response (RIR) $\textbf{h}_q(n,\omega)$  and the STFT signal $s_q(n,\omega)$ of $q^{th}$ source  along the time frame axis for each frequency bin \cite{20}, $n\in {\left\{1,...,N\right\}}$ and $\omega\in {\left\{1,...,\Omega\right\}}$ denote the time frame and frequency bin indices, respectively. $N$ and $\Omega$ denote the maximum time frame and frequency bin indices, respectively. The TF domain microphone array signals can be given as:
\begin{equation}
	\textbf{y}(n,\omega) = \sum^{Q}_{q = 1} \sum_{k^{\prime}=0}^{K} \textbf{h}_q(k^{\prime},\omega)s_q(n-k^{\prime},\omega) + \textbf{n}(n,\omega),
	\label{x}
\end{equation}
where $\textbf{n}(n,\omega)$ denotes the additive noise at the microphone array.
The microphone array signals $\textbf{y}(n,\omega)$ can be reformulated as the sum of the early reverberant component, the late reverberant component and the noise:
\begin{equation}
	\textbf{y}(n,\omega)  =  \textbf{y}_e(n,\omega) + \textbf{y}_l(n,\omega) + \textbf{n}(n,\omega).
	\label{y_e}
\end{equation}
where $\textbf{y}_e(n,\omega) = \sum^{Q}_{q = 1}  \sum_{k^{\prime}=0}^{k_e} \textbf{h}_q(k^{\prime},\omega)s_q(n-k^{\prime},\omega)$  denotes the early reverberant component which contains the direct-path signals and few early reflections, $\textbf{y}_l(n,\omega) = \sum^{Q}_{q = 1} 	\sum_{k^{\prime}=k_e}^{K} \textbf{h}_q(k^{\prime},\omega)s_q(n-k^{\prime},\omega)$ is the late reverberant component which contains all other reflections. $k_e$ is the convolutional order that separates the early reverberant component and the late reverberant component.

When $k^{\prime}=0$, the RIR $\textbf{h}_q(0,\omega)$ is also referred to the steering vector of the $q^{th}$ source from direction $\theta_q$ to the microphone array, defined as $\textbf{a}(\theta_q,\omega) = [ 1,e^{-j\omega \Delta t_1(\theta_q)},...,e^{-j\omega \Delta t_m(\theta_q)},..., e^{-j\omega \Delta t_{M-1}(\theta_q)} ]\in\mathbb{C}^{M\times1}$, where the first microphone is taken as the reference, $\Delta t_m (\theta_q)$ is the time delay from the $m^{th}$ microphone to the reference microphone.

\subsection{MCLP-based methods}
In MCLP-based methods, the dereverberated signals $\textbf{d}(n,\omega)$ can be given as:
\begin{equation}
	\textbf{d}(n,\omega) = \textbf{y}(n,\omega) - \textbf{G}_t^H (\omega) \tilde{\textbf{y}}_t(n,\omega).
	\label{yhat}
\end{equation}
where $\textbf{G}_t(\omega)=[\textbf{g}_{t_1}(\omega),...,\textbf{g}_{t_m}(\omega),...,\textbf{g}_{t_M}(\omega)] \in  \mathbb{C}^{K_tM \times M}  $ is a temporal filter matrix at frequency bin $\omega$ , $\textbf{g}_{t_m}(\omega)$ is the temporal filter vector for the $m^{th}$ microphone. $\tilde{\textbf{y}}_t(n,\omega) = \left[ \textbf{y}^T(n-\Delta_t-1,\omega),...,\textbf{y}^T(n-\Delta_t-K_t,\omega) \right]^T $ is the stacked observation signal matrix, $\Delta_t$ denotes the delay tap index of the beginning frame of the late reverberation component, $K_t$ denotes the prediction order. $(\cdot)^T$ and $(\cdot)^H$ denote the transpose and complex conjugate transpose operator, respectively.
The MCLP-based methods  estimate the filter matrix $\textbf{G}_t(\omega)$ to obtain dereverberated speech signals.


\section{Proposed Algorithm }

The proposed algorithm consists of two parts: dual-path MCLP filters for dereverberation and multi-norm beamforming for denoising.

\subsection{Dual-path MCLP filter branch}

\subsubsection{Dereverberated signals with dual-path filters}

Let us define $\textbf{G}_f (n) = [\textbf{g}_{f_1}(n),...,\textbf{g}_{f_m}(n),...,\textbf{g}_{f_M}(n)] \in  \mathbb{C}^{((2K_f+1)M) \times M}  $ as the frequential filter matrix at time frame $n$, where $\textbf{g}_{f_m}(n)$ is the filter for the $m^{th}$ microphone at time frame $n$, $K_f$ denotes the prediction order in frequential dimension. With both temporal filter matrix in \eqref{yhat} and the proposed frequential filter matrix, the early component in TF domain microphone array signals can be given as:
\begin{equation}
	\textbf{x}(n,\omega)= \textbf{y}(n,\omega) - \textbf{G}_t^H (\omega) \tilde{\textbf{y}}_t(n,\omega)-\textbf{G}_f^H (n) \tilde{\textbf{y}}_f(n,\omega),
	\label{c1}
\end{equation}
where $\tilde{\textbf{y}}_f(n,\omega) = [ \textbf{y}^T(n,\omega-K_f),...,\textbf{y}^T(n,\omega),...,\textbf{y}^T(n,\omega + K_f) ]^T $ is the wide band observation signal matrix at time frame $n$. The dual-path filters in both temporal and frequential dimensions yield more comprehensive modeling of the late reverberation.

\subsubsection{Dual-path MCLP filter}

In the dual-path MCLP filter for dereverberation, we minimize the $l_2$ norm as well as the $l_1$ norm of the dereverberated signals, as the information bearing signals such as speech signals exhibit sparsity in their STFT coefficients:
\begin{align}
	\left\lbrace  \hat{\textbf{G}}_t(\omega), \hat{\textbf{G}}_f(n) \right\rbrace=\underset{\textbf{G}_t,\textbf{G}_f}{{\arg\min}}\sum_{n=1}^{N} \sum_{\omega=1}^{\Omega} &( \Vert \textbf{x}(n,\omega)\Vert^2_2	\nonumber  \\
	& +\lambda_{\textbf{z}}\Vert \textbf{x}(n,\omega) \Vert_1 ) ,
	\label{dp mclp}
\end{align}
where $\textbf{x}(n,\omega)$ is given in \eqref{c1}, $\lambda_{\textbf{z}}$ is sparsity penalization parameter, $\hat{\textbf{G}}_t(\omega)$, $\hat{\textbf{G}}_f(n)$ are the estimated $\textbf{G}_t$ and $\textbf{G}_f$, respectively. For the sake of simplicity, $\omega$ and $n$ in $\textbf{G}_t(\omega)$, $\textbf{G}_f(n)$,   $\tilde{\textbf{y}}_t(n,\omega)$, and $\tilde{\textbf{y}}_f(n,\omega)$ are omitted in the following of this paper. The problem \eqref{dp mclp} can be solved via Proximal Alternating Linearized Minimization (PALM) \cite{21} method. By introducing an extra variable $\textbf{z}(n,\omega)=\textbf{x}(n,\omega)$, the augmented Lagrangian of \eqref{dp mclp} can be given as:
\begin{align}
	&  \mathcal{L}(\textbf{G}_t,\textbf{G}_f, \textbf{z}(n,\omega), \boldsymbol{\eta}) =  \sum_{n=1}^{N} \sum_{\omega=1}^{\Omega} (\Vert\textbf{x}(n,\omega)
	\Vert^2_2+\lambda_{\textbf{z}} \Vert   \textbf{z}(n,\omega) \Vert_1\nonumber   \\
	 & +\mathcal{R}e\lbrace \boldsymbol{\eta}^H(\textbf{x}(n,\omega)-\textbf{z}(n,\omega) )\rbrace 
	+  \frac{1}{2\rho_{\textbf{G}}}\Vert\textbf{x}(n,\omega) -\textbf{z}(n,\omega)  \Vert^2_2 ),
	\label{LM dp mclp}
\end{align}
where $\mathcal{R}e \left\lbrace \cdot \right\rbrace$ is the real part operator, $\rho_{\textbf{G}}$ is the penalization parameter of the convex term and $\boldsymbol{\eta}$ is the Lagrange multiplier vector. Utilising the method in \cite{22}, the solutions for $\textbf{G}_t$, $\textbf{G}_f$, $\textbf{z}(n,\omega)$, and $\boldsymbol{\eta}$ in the $(l+1)^{th}$ iteration are given as:
\begin{align}
	& \textbf{G}_t^{(l+1)} = \left( \sum_{n=1}^{N} \left( 1+\frac{1}{2\rho_{\textbf{G}}} \right)  \tilde{\textbf{y}}_t \tilde{\textbf{y}}_t^H \right)^{-1} 
	\nonumber  \\ 
	& \hspace{1.2cm}\cdot \sum_{n=1}^{N} \left( 1+\frac{1}{2\rho_{\textbf{G}}} \right) \tilde{\textbf{y}}_t \textbf{y}^H(n,\omega) + \frac{1}{2}\tilde{\textbf{y}}_t\boldsymbol{\eta}^{(l)H}   
	\nonumber  \\
	& \hspace{1.2cm}-\left( 1+\frac{1}{2\rho_{\textbf{G}}}  \right)\tilde{\textbf{y}}_t\tilde{\textbf{y}}_f^H\textbf{G}_f^{(l)}  - \frac{1}{2\rho_{\textbf{G}}} \tilde{\textbf{y}}_t\textbf{z}^{(l)H}(n,\omega) ,
	\label{SolveG_t}
\end{align}
and:
\begin{align}
	& \textbf{G}_f^{(l+1)} = \left( \sum_{\omega=1}^{\Omega} \left( 1+\frac{1}{2\rho_{\textbf{G}}} \right)  \tilde{\textbf{y}}_f \tilde{\textbf{y}}_f^H \right)^{-1} 
	\nonumber  \\ 
	&  \hspace{1.2cm}\cdot \sum_{\omega=1}^{\Omega}  \left( 1+\frac{1}{2\rho_{\textbf{G}}}  \right) \tilde{\textbf{y}}_f\textbf{y}^H(n,\omega) + \frac{1}{2}\tilde{\textbf{y}}_f\boldsymbol{\eta}^{(l)H}   
	\nonumber  \\
	& \hspace{1.2cm}-\left( 1+\frac{1}{2\rho_{\textbf{G}}}  \right)\tilde{\textbf{y}}_f\tilde{\textbf{y}}_t^H\textbf{G}_t^{(l+1)}  - \frac{1}{2\rho_{\textbf{G}}} \tilde{\textbf{y}}_f 	\textbf{z}^{(l)H}(n,\omega) , 
	\label{SolveG_f}
\end{align}
and:
\begin{align}
&	\textbf{z}^{(l+1)} (n,\omega)
	=  \nonumber   \\
&	\mathcal{S}_{\lambda_{\textbf{z}}/\mu_{\textbf{z}}} \left( \textbf{z}^{(l)}(n,\omega) - \frac{1}{\mu_{\textbf{z}}} \nabla_{\textbf{z}}V(\textbf{G}_t^{(l+1)},\textbf{G}_f^{(l+1)}, \textbf{z}^{(l)}(n,\omega), \boldsymbol{\eta}^{(l)})  \right),
	\label{Solve_Z}
\end{align}
where $V(\textbf{G}_t^{(l+1)},\textbf{G}_f^{(l+1)},\textbf{z}^{(l)}(n,\omega), \boldsymbol{\eta}^{(l)}) = \mathcal{R}e \lbrace\boldsymbol{\eta}^{(l)H}(\textbf{y}(n,\omega)-\textbf{G}_t^{(l+1)H}\tilde{\textbf{y}}_t-\textbf{G}_f^{(l+1)H}\tilde{\textbf{y}}_f-\textbf{z}^{(l)}(n,\omega))\rbrace+\frac{1}{2\rho_{\textbf{G}}}  \Vert \textbf{y}(n,\omega) - \textbf{G}_t^{(l+1)H}\tilde{\textbf{y}}_t -\textbf{G}_f^{(l+1)H}\tilde{\textbf{y}}_f- \textbf{z}^{(l)}(n,\omega)\Vert^2_2$, $\mu_{\textbf{z}}$ is thresholding parameter and $\mathcal{S}_{\lambda_{\textbf{z}}/\mu_{\textbf{z}}}(\textbf{v})$ is the soft thresholding operator \cite{23} of the vector $\textbf{v}$ such that:
\[
S_{\lambda_{\textbf{z}}/\mu_{\textbf{z}}}(\mathbf{v}) = 
\begin{cases}
	\mathbf{v} - \lambda_{\textbf{z}}/\mu_{\textbf{z}}, & \text{if } \mathbf{v} \geq \lambda_{\textbf{z}}/\mu_{\textbf{z}}, \\
	\mathbf{v} + \lambda_{\textbf{z}}/\mu_{\textbf{z}}, & \text{if } \mathbf{v} \leq -\lambda_{\textbf{z}}/\mu_{\textbf{z}}, \\
	0, & \text{otherwise}
\end{cases}
\]
and:
\begin{align}
	&\boldsymbol{\eta}^{(l+1)}=
	\nonumber   \\
	& \boldsymbol{\eta}^{(l)}+\gamma(\textbf{y}(n,\omega) - \textbf{G}_t^{{(l+1)}H}\tilde{\textbf{y}}_t -\textbf{G}_f^{{(l+1)}H}\tilde{\textbf{y}}_f- \textbf{z}^{(l+1)}(n,\omega)),
	\label{Lagragian2sGc}
\end{align}
where $\gamma$ is the step size parameter.

With $\hat{\textbf{G}}_t$ and $\hat{\textbf{G}}_f$ estimated in \eqref{dp mclp}, the estimated early component of microphone signals can be given as:
\begin{equation}
	\hat{\textbf{x}}(n,\omega)= \textbf{y}(n,\omega) - \hat{\textbf{G}}_t^H  \tilde{\textbf{y}}_t - \hat{\textbf{G}}_f^H  \tilde{\textbf{y}}_f.
	\label{xhat}
\end{equation}
\subsection{Multi-norm beamforming branch}

The multi-norm beamforming is applied to the signals dereverberated by the dual-path MCLP filters, namely $\hat{\textbf{x}}(n, \omega)$ in \eqref{xhat}.  The power as well as the $l_1$ norm of the denoised signals are minimized while the signals from the target direction of arrival (DOA) are preserved:
\begin{align}
	&  \hat{\textbf{w}} =   \underset{\textbf{w}}{{\arg\text{min}}}  \sum_{n=1}^{N} \left( \Vert \textbf{w}^H\hat{\textbf{x}}(n, \omega)\Vert^2_2  + \lambda_{\mathbf{w}} \Vert \textbf{w}^H\hat{\textbf{x}}(n, \omega) \Vert_1 \right) , \nonumber  \\
	& 
		\quad  \text{s.t.} \quad \textbf{w}^H\textbf{a}(\theta_s)=1,    	
	\label{beamforming}
\end{align}
where $\textbf{w}$ is the beamforming filter vector, $\hat{\textbf{w}}$ is the estimated $\textbf{w}$ and $\lambda_{\textbf{w}}$ is the sparsity penalization parameter.

The problem \eqref{beamforming} can  be solved via Alternating Direction Method of Multipliers (ADMM) method \cite{24}. Introducing an extra variable $z_\textbf{w}(n) = \textbf{w}^H\hat{\textbf{x}}(n,\omega)$, the augmented Lagrangian of \eqref{beamforming} can be given as:
\begin{align}
	&  \mathcal{L}(\textbf{w} , z_{\textbf{w}}(n) , \eta_\textbf{w}) =  \sum_{n=1}^{N}\left(  \Vert   \textbf{w}^H \hat{\textbf{x}}(n, \omega)
	\Vert^2_2 + \lambda_{\mathbf{w}} \left\lVert z_{\mathbf{w}}(n) \right\rVert_1 \right. \nonumber   \\
	& \left. +\mathcal{R}e\lbrace  \eta_\textbf{w}^H (\textbf{w}^H\hat{\textbf{x}}(n, \omega)-  z_{\textbf{w}}(n)) \rbrace  \right. \nonumber  \\
	& \left. +\frac{1}{2\rho_{\textbf{w}}}\Vert   \textbf{w}^H \hat{\textbf{x}}(n, \omega)-z_{\textbf{w}}(n)
	\Vert^2_2 \right),  \quad \text{s.t. }   \textbf{w}^H\textbf{a}(\theta_s) =1.
\label{Lagragian2sG_W}
\end{align}
where $\eta_\textbf{w}$ is Lagrange multiplier. The problem \eqref{Lagragian2sG_W} can be solved via some iterative steps similar to problem \eqref{LM dp mclp}. In the $(l+1)^{th}$ iteration, the details for solving $\textbf{w}^{(l+1)}$ will be given in Appendix. $z_{\textbf{w}}^{(l+1)}(n) $ can be given as:
\begin{align}
&	z_{\textbf{w}}^{(l+1)}(n) \nonumber  \\
&	= \mathcal{S}_{\lambda_{\textbf{w}}/\mu_{\textbf{w}}} \left( z_{\textbf{w}}^{(l)}(n) - \frac{1}{\mu_{\textbf{w}}} \nabla_{\textbf{z}_{\textbf{w}}}V(\textbf{w}^{(l+1)},z_{\textbf{w}}^{(l)}(n), \eta_{\textbf{w}}^{(l)})  \right)
	\label{SolveZ},
\end{align}
where $V(\textbf{w}^{(l+1)},z_{\textbf{w}}^{(l)}(n), \eta_{\textbf{w}}^{(l)}) =   \mathcal{R}e \lbrace  \eta_\textbf{w}^{(l)H} (\textbf{w}^{(l+1)H} 
\hat{\textbf{x}}(n,\omega)$ $z_{\textbf{w}}^{(l)}(n)) \rbrace+\frac{1}{2\rho_{\textbf{w}}}\Vert   \textbf{w}^{(l+1)H} 
\hat{\textbf{x}}(n, \omega)-  z_{\textbf{w}}^{(l)}(n)
\Vert^2_2$,  $\mu_{\textbf{w}}$ is thresholding parameter and $\mathcal{S}_{\lambda_{\textbf{w}}/\mu_{\textbf{w}}}(\cdot)$ is the soft thresholding operator \cite{23}, $\eta_{\textbf{w}}^{(l+1)}$ can be given as:
\begin{equation}
	\eta_{\textbf{w}}^{(l+1)}=\eta_{\textbf{w}}^{(l)}+\gamma_{\textbf{w}}(\textbf{w}^{(l+1)H}\hat{\textbf{x}}(n,\omega)- z_{\textbf{w}}^{(l+1)}(n) ).
\end{equation}
where $\gamma_{\textbf{w}}$ is the step size parameter.

With the estimation result $\hat{\textbf{w}}$ obtained by \eqref{beamforming}, the enhanced speech signals can be given as:
\begin{equation}
\hat{s}(n,\omega) = \hat{\textbf{w}}^H \hat{\textbf{x}}(n,\omega).
\label{shat}
\end{equation}

\section{Proposed Prediction Order Selection Method}
\begin{figure}[t]
	\centering
	\begin{minipage}{\linewidth}
		\centering
		\subfigure[Pearson correlation coefficients]{\includegraphics[width=0.48\linewidth, height=4cm, keepaspectratio]{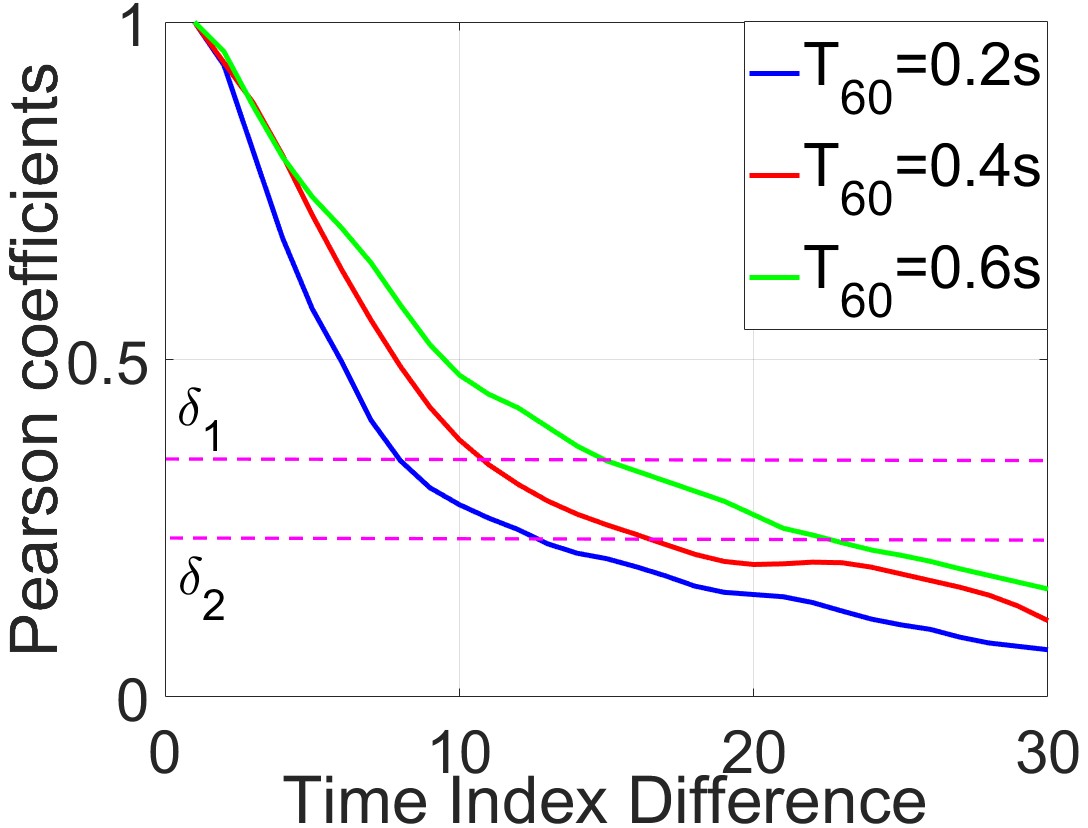}}
		\hfill
		\subfigure[PESQ with $T_{60}=0.2$s]{\includegraphics[width=0.48\linewidth, height=4cm, keepaspectratio]{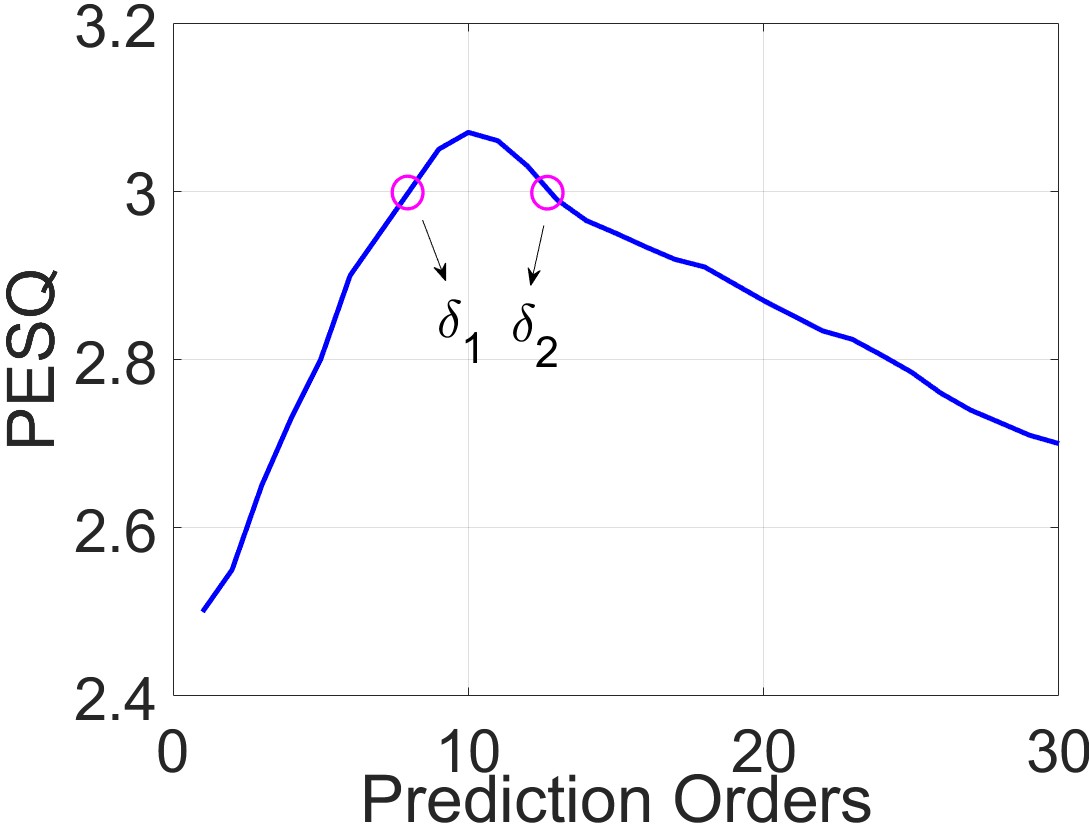}}
		\subfigure[PESQ with $T_{60}=0.4$s]{\includegraphics[width=0.48\linewidth, height=4cm, keepaspectratio]{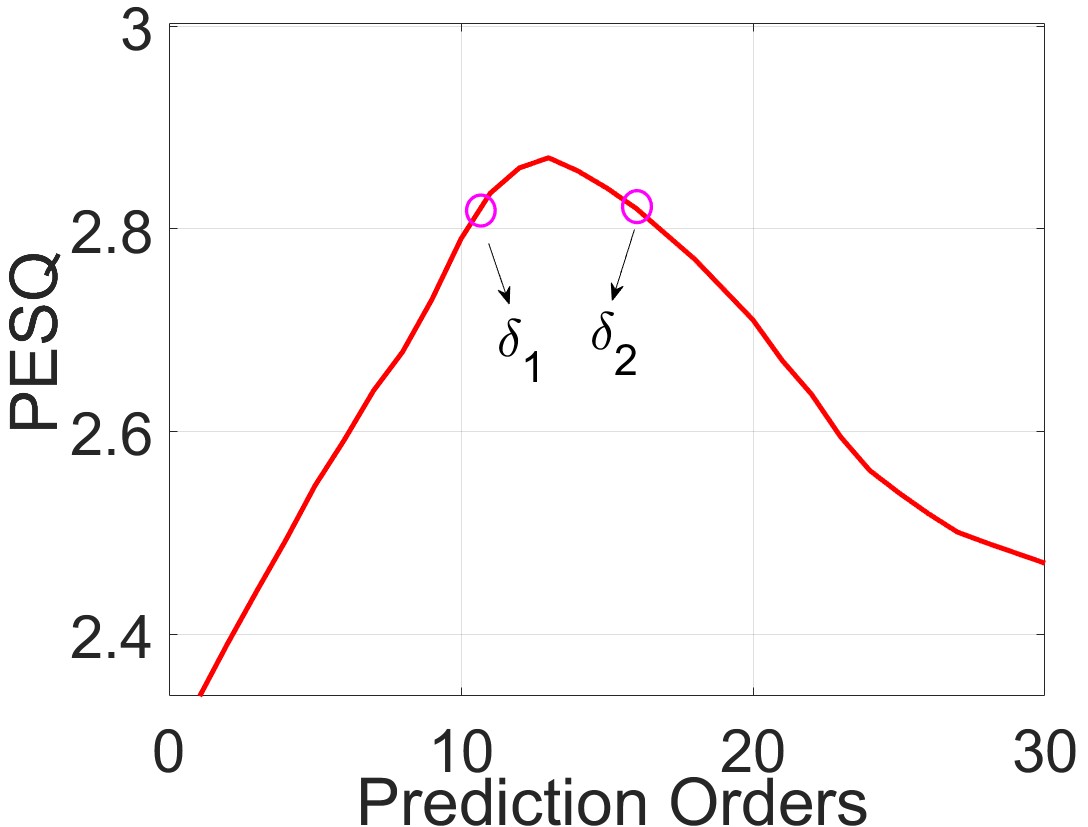}}
		\hfill
		\subfigure[PESQ with $T_{60}=0.6$s]{\includegraphics[width=0.48\linewidth, height=4cm, keepaspectratio]{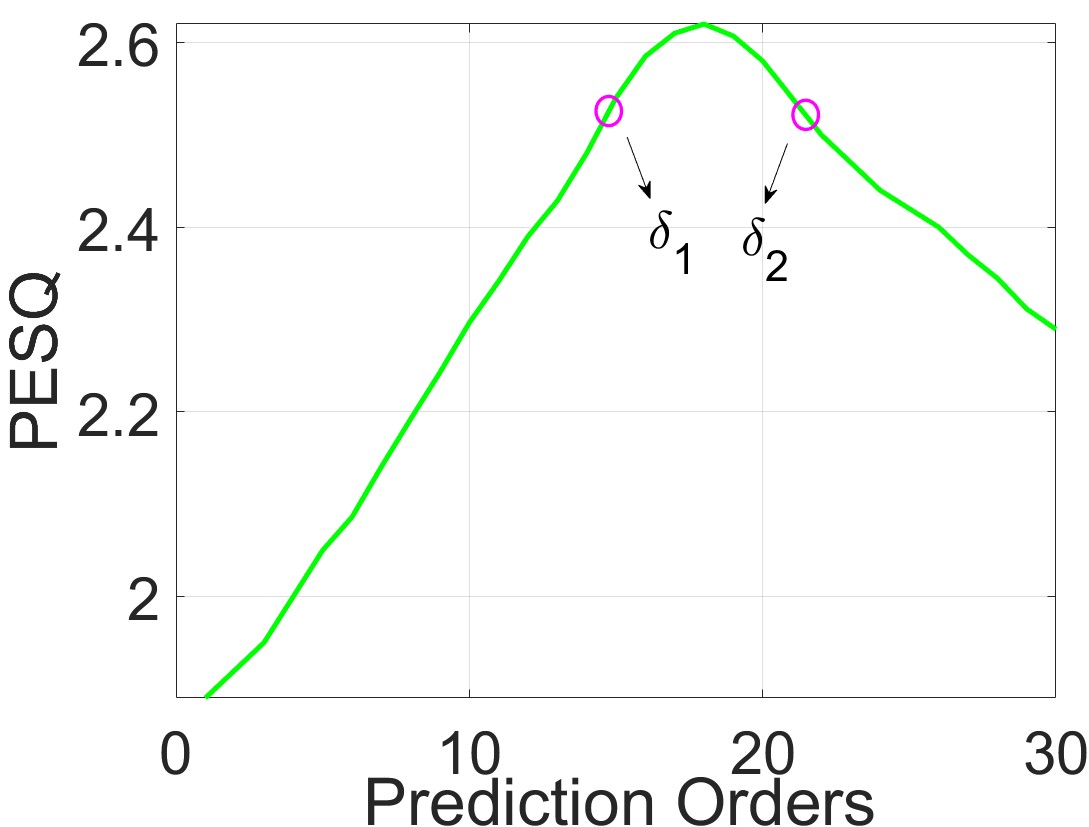}}
	\end{minipage}
	\caption{Pearson correlation coefficients and PESQ with different $T_{60}$ values.}
	\label{Peareson}
\end{figure}
In this section, the prediction order selection method is present. Firstly, let us define $y_{1_t}^i(t)$, $t = 0,...,T$ as the $i^{th}$ individual sample of the temporal signals at the reference microphone at time index $t$ in the Monte Carlo experiments. The Pearson correlation coefficients between $y_{1_t}^i(0)$ and $y_{1_t}^i(t)$ can be given as:
\begin{align}
&	\varrho_{y_{1_t}}(0,t)=   \nonumber  \\
&	\frac{\sum_{i=1}^{I}(y_{1_t}^i(0)-\bar{y}_{1_t}(0))(y_{1_t}^i(t)-\bar{y}_{1_t}(t))}{\sqrt{\sum_{i=1}^{I}(y_{1_t}^i(0)-\bar{y}_{1_t}(0))^2\sum_{i=1}^{I}(y_{1_t}^i(t)-\bar{y}_{1_t}(t))^2}},
\label{pearson}
\end{align}
where $I$ denotes the total sample number, $(\bar{\cdot})$ denotes the mean value operator. In Fig. \ref{Peareson}(a) the Pearson correlation coefficients $\varrho_{y_{1_t}}(0,t)$ in \eqref{pearson} are presented as a function of time lag $t$ under different $T_{60}$ with $I=50$, namely $50$ Monte Carlo experiments.   It can be observed that, due to the appearance of the late reverberation, $\varrho_{y_{1_t}}(0,t)$ decreases gently as the time lag increases. 

In our experiments, we have found that, for different $T_{60}$ configurations, a single threshold $\delta$ can be selected within the range $\delta_1\leq \delta\leq\delta_2$, such that the corresponding time lag is also an approximate optimal prediction order of the temporal filters in our proposed MCLP based method. This can be illustrated in Fig.\ref{Peareson} (b)-(d), where the Perceptual Evaluation of Speech Quality (PESQ) of our proposed method is plotted as a function of the temporal filter prediction order $K_t$ under different $T_{60}$ values, while the frequential filter prediction order $K_f$ is fixed. The prediction orders corresponding to the rose-red circles of PESQ values in Fig. \ref{Peareson}(b)-(d) are equal to the time lags corresponding to $\delta_1$ and $\delta_2$ in Fig. \ref{Peareson}(a). Hence, in our following work, the optimal prediction order in the time filters is selected as:
\begin{align}
	K_t = 1/2(K_{\delta_1} + K_{\delta_2}),
\end{align}
where  $K_{\delta_1}$ and $K_{\delta_2}$ are the time lags corresponding to $\delta_1$ and $\delta_2$ in Fig. \ref{Peareson}(a), respectively. In this manner, an approximate optimal prediction order for different $T_{60}$ values can be determined by calculating the Pearson correlation coefficients between the temporal signals at different time lags, which is efficient for real-world applications. This optimal prediction order selection approach can be applied to other MCLP-based method as well.

The optimal prediction order in the frequential filters in our proposed method can be determined in a similar way.

\section{Simulation Experiments}
In this section, numerical simulations are presented to illustrate the validity of the proposed method. 
Source signals are speeches from the TIMIT database \cite{25} sampled at $16$ kHz.
A uniform linear microphone array (ULA) composed of $8$ microphones with the inter-element space equal to $0.03$ m is utilised. 
The RIRs are generated using the image method \cite{26} in a room with size of $6$ m $\times$ $6$ m $\times$ $3$ m and the additive noises are Gaussian white.
The Hanning windows with the length of $512$ samples are used for  signal analysis and synthesis with hop size of half-length. $100$ Monte Carlo simulations are conducted for each configuration.

In all the experiments, our proposed method is compared with the baseline methods including the GWPE method \cite{9} (legend: “GWPE”) , the cascade of  GWPE and MVDR method \cite{12} (legend: “GWPE+MVDR”) and the Weighted Power minimization Distortionless response (WPD) beamforming \cite{14} (legend: “WPD”) method. The PESQ and Scale-Invariant Signal-to-Noise Ratio (SI-SNR) are used as the evaluation metrics.

\begin{figure}[t]
	\centering
	\begin{minipage}{\linewidth}
		\centering
		\subfigure[PESQ]{\includegraphics[width=0.49\linewidth, height=4cm, keepaspectratio]{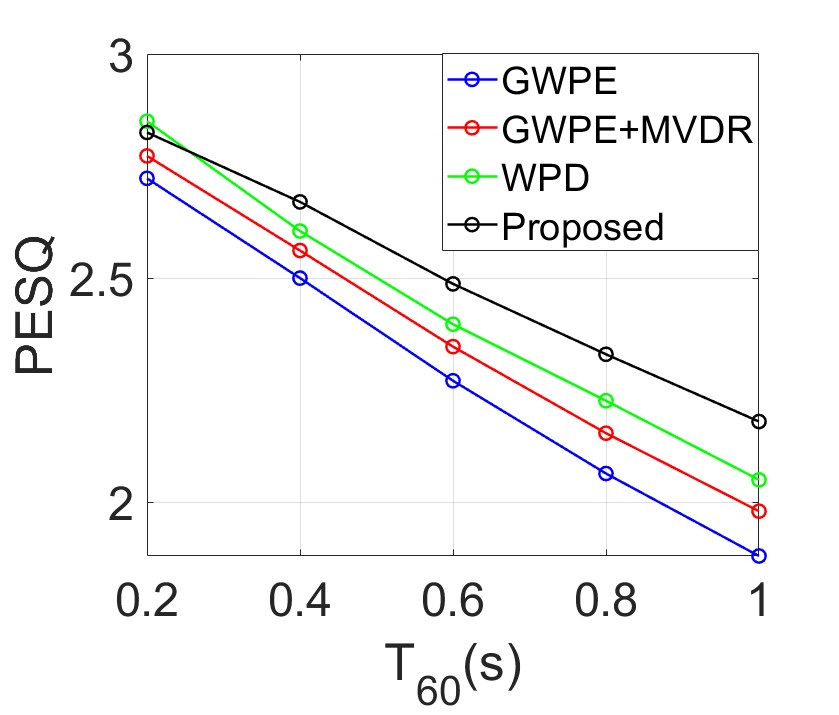}} \label{different T601}
		\hfill
		\subfigure[SI-SNR]{\includegraphics[width=0.49\linewidth, height=4cm, keepaspectratio]{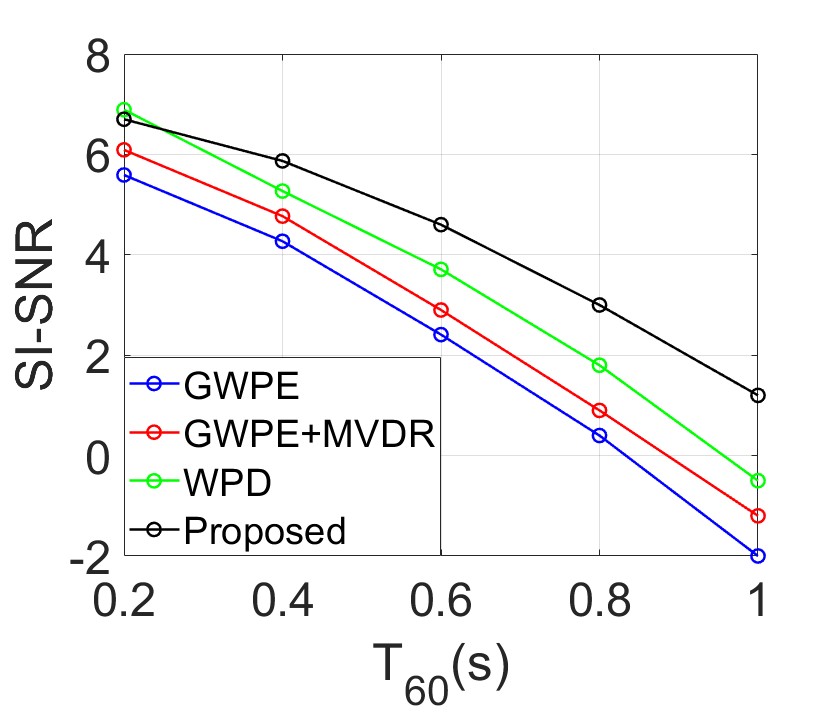}} \label{different T602}
		\subfigure[PESQ]{\includegraphics[width=0.49\linewidth, height=4cm, keepaspectratio]{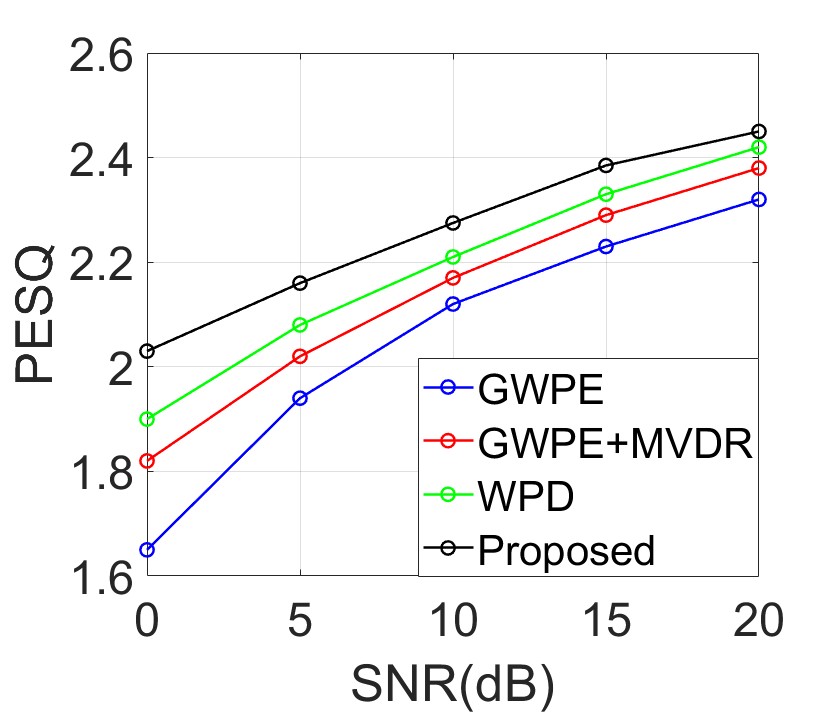}}  \label{different SNR1}
		\hfill
		\subfigure[SI-SNR]{\includegraphics[width=0.49\linewidth, height=4cm, keepaspectratio]{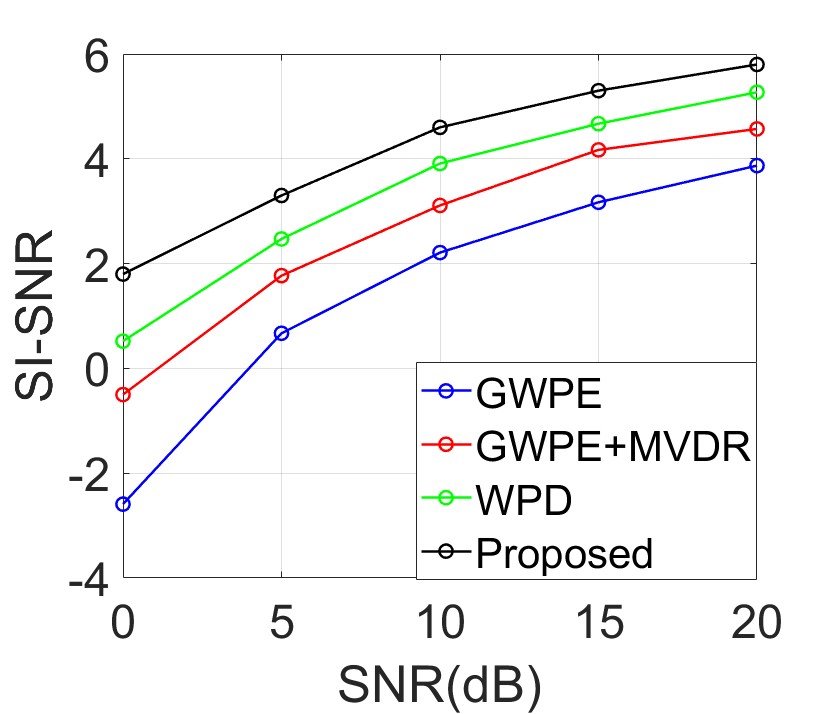}}  \label{different SNR2}
	\end{minipage}
	\caption{Performance metrics under different configurations.}
	\label{metrics}
\end{figure}

In Fig. \ref{metrics}(a) and \ref{metrics}(b), PESQ and SI-SNR obtained by the baseline methods and the proposed method are plotted as a function of $T_{60}$, with SNR set to $25$ dB. $K_t=\{10, 14, 18, 22, 24\}$  and $K_f=\{2, 4, 6, 8, 10\} $ for $T_{60}=\{0.2 s, 0.4 s, 0.6 s, 0.8 s, 1 s\}$. It can be seen that, with the dual-path MCLP filters which can more comprehensively removing the late reverberation,  the advantage of the proposed method becomes more significant in high reverberation cases. When $T_{60}$ is small enough ($200$ ms in this experiment), the performance of the proposed method is slightly worse than that of the baseline method WPD. One plausible explanation is that  temporal filters alone are sufficient for removing the late reverberation when $T_{60}$ is sufficiently small. A trade-off between dual-path MCLP filters or temporal-only MCLP filters should be considered based on varying $T_{60}$ in our proposed method.

Fig. \ref{metrics}(c) and \ref{metrics}(d) show PESQ and SI-SNR as a function of SNR, with $T_{60}$ set to $0.3 s$ . $K_t=12$ and $K_f=3$. It can be observed that, 
as a joint optimization method, WPD outperforms the cascaded GWPE and MVDR method ("GWPE+MVDR"). 
However, due to the $l_1$ norm constraint on the denoised signals, our proposed method exhibits a more powerful denoising capability and outperforms all the baseline methods across all the SNRs.

\section{Conclusion}

In this paper, we propose a speech enhancement method using dual-path MCLP filters and multi-norm beamforming. The proposed method demonstrates superior performance in both dereverberation and denoising compared to the baseline methods, particularly in high reverberation scenarios. In addition, we have proposed an efficient method for selecting  optimal prediction orders in both temporal and frequential filters for the proposed method, which can also be applied to other MCLP-based methods.  Simulation results validate the effectiveness of our proposed method.

\section{Appendix}
To solve $\textbf{w}^{(l+1)}$, a new augmented Lagrangian can be derived as:
\begin{align}
	&  \mathcal{L}(\textbf{w}^{(l+1)} , \eta_1 ) =  \sum_{n=1}^{N}(\Vert   \textbf{w}^{(l+1)H}\hat{\textbf{x}}(n,\omega)
	\Vert^2_2 +\mathcal{R}e\lbrace  \eta_\textbf{w}^{(l)H} (\textbf{w}^{(l+1)H} \nonumber   \\
	& \hat{\textbf{x}}(n,\omega)-  z_{\textbf{w}}^{(l)}(n,\omega)) \rbrace+\frac{1}{2\rho_{\textbf{w}}}\Vert   \textbf{w}^{(l+1)H}\hat{\textbf{x}}(n,\omega)- z_{\textbf{w}}^{(l)}(n,\omega)
	\Vert^2_2 ) + \nonumber   \\
	&\mathcal{R}e\lbrace  \eta_1^H (\mathbf{w}^{(l+1)H}\mathbf{a}(\theta_s) -1) \rbrace +\frac{1}{2\rho_1}\Vert   \textbf{w}^{(l+1)H} \mathbf{a}(\theta_s) -1
	\Vert^2_2	,
	\label{LagraW}
\end{align}
The problem \eqref{LagraW} can be solved via several iterative steps. In the $(j+1)^{th}$ iteration, $\textbf{w}$ can be given as:
\begin{align}
	& \textbf{w}^{(l+1,j+1)} = \left(\sum_{n=1}^{N}(1+\frac{1}{2\rho_{\textbf{w}}})\hat{\textbf{x}}(n,\omega)\hat{\textbf{x}}^H(n,\omega)+\frac{\textbf{a}(\theta_s)\textbf{a}^H(\theta_s)}{2\rho_1}\right)^{-1} 
	\nonumber  \\ 
	&\left(\sum_{n=1}^{N} \hat{\textbf{x}}(n,\omega)(\frac{1}{2\rho_{\textbf{w}}}z_{\textbf{w}}^{(l)H}(n,\omega)-\frac{1}{2}\eta_{\textbf{w}}^{(l)H})+\textbf{a}(\theta_s)(\frac{1}{2\rho_1}  -\frac{1}{2}\eta_1^{(j)H})\right),
\end{align}
where $\eta_1^{(j+1)}$ can be given as:
\begin{align}
\eta_1^{(j+1)}=\eta_1^{(j)}+\gamma_1(\textbf{w}^{(l+1,j+1)H} \textbf{a}(\theta_s)- 1).
\end{align}

\section{Acknowledgements}
This work was supported by the National Natural Science Foundation of China (Grant No.62101013).

\bibliographystyle{IEEEtran}
\bibliography{mybib}

\end{document}